\documentclass[achemso,reprint]{revtex4-1}

\usepackage[T1]{fontenc}
\usepackage{color}
\usepackage{amsmath}
\usepackage{amssymb}
\usepackage{graphicx}
\usepackage{esint}
\usepackage{lineno}
\usepackage{lipsum}

\definecolor{samblue}{rgb}{0.0,0.2,0.5}
\definecolor{samgreen}{rgb}{0,0.5,0.0}
\definecolor{samred}{rgb}{0.66,0,0}

\usepackage[normalem]{ulem}

\makeatletter

\renewcommand{\fnum@figure}{\textbf{Figure~\thefigure}}
\usepackage{units}
\usepackage{lipsum}
\usepackage{babel}

\makeatother

\usepackage{babel}
\begin{document}

\title{Bosonic Delocalization of Dipolar Moir\'e Excitons}

\author{Samuel Brem}
\email{brem@uni-marburg.de}
\author{Ermin Malic}
\affiliation{Department of Physics, Philipps University, 35037 Marburg, Germany}

\begin{abstract}
In superlattices of twisted semiconductor monolayers, tunable moir\'e potentials emerge, trapping excitons into periodic arrays. In particular, spatially separated interlayer excitons are subject to a deep potential landscape and they exhibit a permanent dipole providing a unique opportunity to study interacting bosonic lattices. Recent experiments have demonstrated density-dependent transport properties of moir\'e excitons, which could play a key role for technological applications. However, the intriguing interplay between exciton-exciton interactions and moir\'e trapping has not been well understood yet. In this work, we develop a microscopic theory of interacting excitons in external potentials allowing us to tackle this highly challenging problem. We find that interactions between moir\'e excitons lead to a delocalization at intermediate densities and we show how this transition can be tuned via twist angle and temperature. The delocalization is accompanied by a modification of optical moir\'e resonances, which gradually merge into a single free exciton peak. The predicted density-tunability of the supercell hopping can be utilized to control the energy transport in moir\'e materials.
\end{abstract}

\maketitle
\begin{figure}[t!]
\includegraphics[width=0.95\columnwidth]{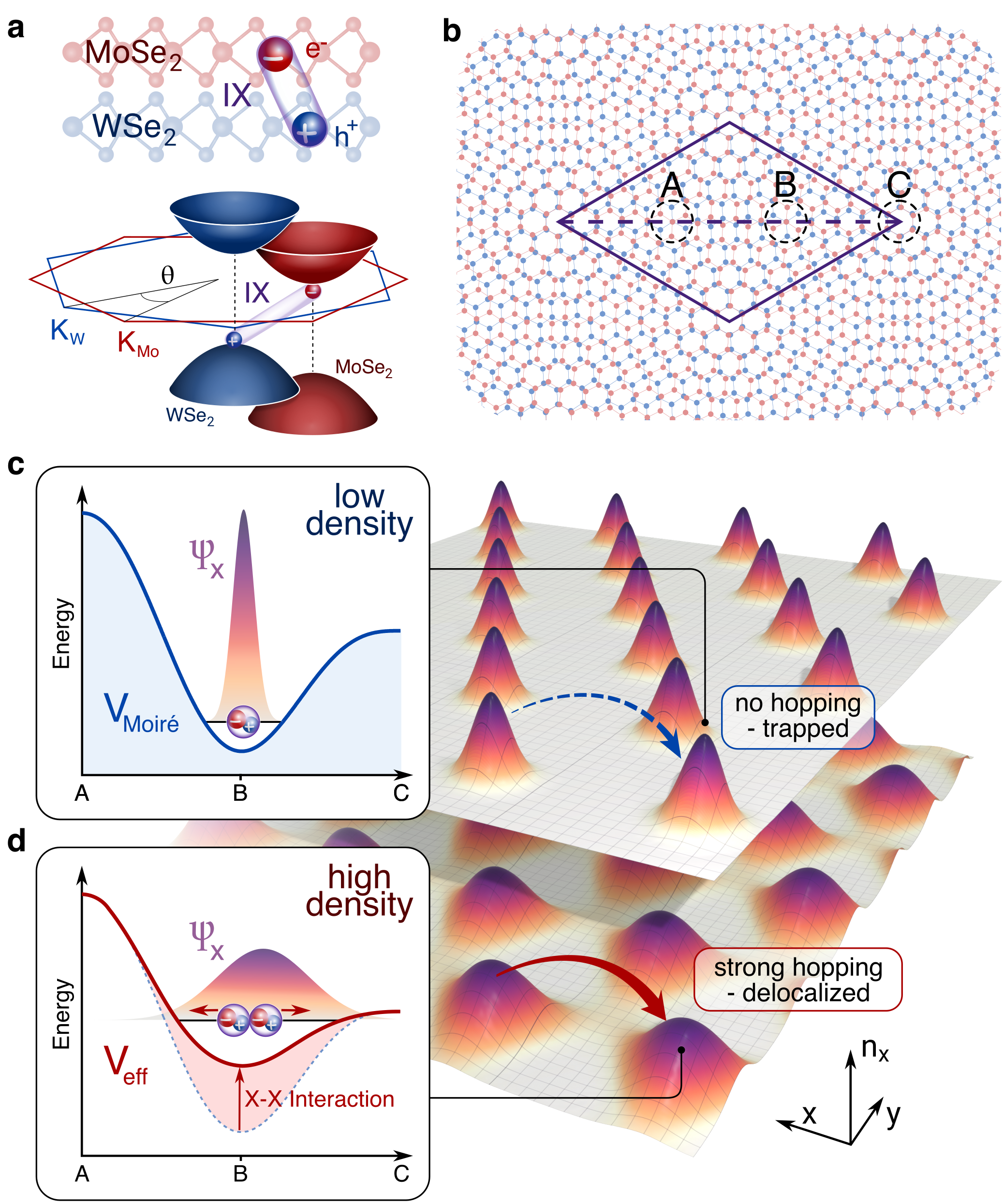}
\caption{{\bf{Interlayer moir\'e excitons.}} {\bf{a}}, In MoSe$_2$/WSe$_2$ heterobilayers, the charge-separated interlayer exciton at the K point (IX) is the energetically most favourable electron-hole configuration. {\bf{b}}, When the two monolayers are slightly twisted a long-range moir\'e pattern is formed with high-symmetry sites (A,B,C). The spatial variation of the atomic alignment leads to periodic changes in the exciton energy. {\bf{c}}, These moir\'e potentials ($V_\text{Moir\'e}$) capture excitons within their minima creating arrays of localized excitons. {\bf{d}}, For large  densities  the inter-excitonic repulsion gives rise to a decrease of the effective potential and a change of the exciton wave function with far-reaching consequences for optical properties and exciton transport.}
\label{fig:scheme} 
\end{figure}

Atomically engineered quantum materials provide an unprecedented platform to experimentally study exotic many-body correlations \cite{wilson2021excitons, andrei2021marvels,mak2022semiconductor}. Recently, artificially stacked superlattices of different monolayer semiconductors, in particular van der Waals heterostructures of transition metal dichalogenides (TMDCs), have been demonstrated to host correlated electron states \cite{shimazaki2020strongly, wang2020correlated}, such as Mott insulators \cite{tang2020simulation, regan2020mott,li2021continuous} and generalized Wigner crystals \cite{wu2018hubbard, zhou2021bilayer, pan2020quantum, huang2021correlated}. Moreover, the non-equilibrium dynamics of these atomically thin semiconductors is governed by tightly bound excitons \cite{chernikov2014exciton,he2014tightly, wang2018colloquium, perea2022exciton}, whose interaction with the stacking tunable superlattice potential opens up a fascinating opportunity to study correlated bosonic states \cite{goral2002quantum,greiner2002quantum}.  
When e.g. a molybdenum diselenide (MoSe$_2$) monolayer  is stacked on top of a tungsten diselenide (WSe$_2$) monolayer, their type II band alignment gives rise to a spatially indirect band gap at the K points of the Brillouin zones \cite{lu2019modulated}. This means that the interlayer exciton \cite{rivera2015observation, miller2017long} with electron and hole residing in different layers is the energetically most favourable state (Fig. \ref{fig:scheme}a). Consequently, after optical excitation of the monolayer intralayer exciton, electrons and holes are spatially separated on a femtosecond timescale \cite{jin2018ultrafast, merkl2019ultrafast, schmitt2022formation, meneghini2022ultrafast} forming a population of long-lived interlayer excitons. These have highly interesting properties arising from their strong dipole-type interaction as well as the moir\'e pattern of the superlattice.

\begin{figure*}[t!]
\includegraphics[width=\textwidth]{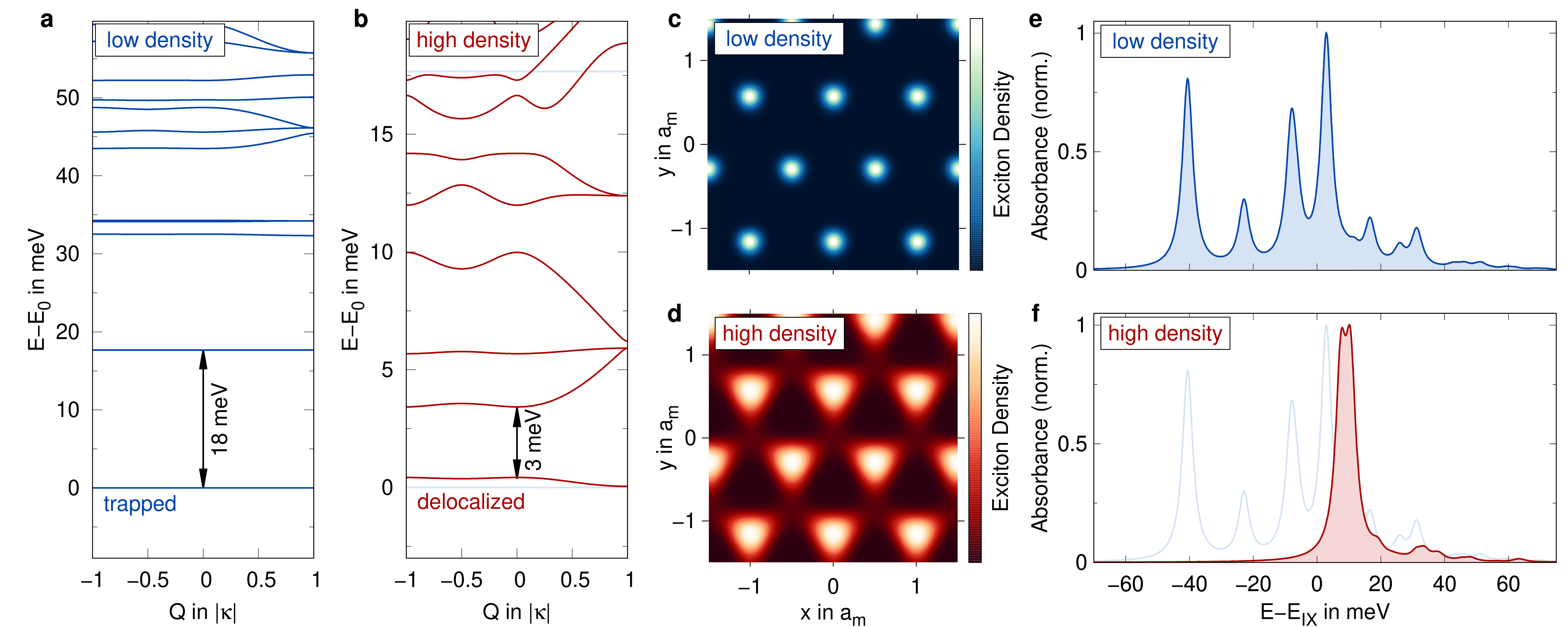}
\caption{{\bf{Density-dependent moir\'e exciton properties.}} {\bf{a-b}}, The mini-band structure of interlayer excitons in hBN encapsulated MoSe$_2$/WSe$_2$ with $1^\circ$ twist angle and cryogenic temperature ($T=4K$) is displayed on a straight path through the hexagonal mini-Brillouin zone comparing the case of a low ({\bf{a}}, $n_X=10^8$cm$^{-2}$) and a high density ({\bf{b}}, $n_X=10^{12}$cm$^{-2}$). Energies are plotted with respect to the lowest state $E_0$. The widely separated flat bands of localized states at low density become continuous bands with smaller gaps reflecting delocalization at high density. {\bf{c-d}} Real space maps of the self-consistent exciton density obtained for the same systems as in {\bf{a}} and {\bf{b}}. {\bf{e-f}} The corresponding absorbance spectra plotted with respect to the energy of the unperturbed interlayer exciton $E_\text{IX}$ (i.e. without the moir\'e potential). As result of the interaction-induced delocalization, the series of moir\'e resonances vanishes and we find a single dominant feature closely resembling the unperturbed exciton peak.}
\label{fig:loc-deloc} 
\end{figure*}  
When two monolayers are stacked with a small twist angle, a long-range moir\'e pattern is formed \cite{shabani2021deep}, where the local atomic registry is changing periodically (Fig. \ref{fig:scheme}b). The interlayer bandgap is, however, strongly affected by the atomic alignment \cite{lu2019modulated,zhang2017interlayer}, such that interlayer excitons are subject to a spatially periodic energy landscape. At small twist angles, i.e. long superlattice periods, these moir\'e potentials lead to arrays of localized exciton states \cite{yu2017moire,brem2020tunable,huang2022excitons} (Fig.\ref{fig:scheme}c). Moir\'e trapped excitons exhibit a series of experimentally confirmed emission features \cite{tran2019evidence,jin2019observation} reflecting the energy spectrum of localized orbitals. Such a network of localized excitons represents an experimentally tunable realization of the Bose-Hubbard model and could enable the study of bosonic supersolid and superfluid states \cite{lagoin2021key,gotting2022moire}. 

However, for an experimental realization of correlated states of interlayer moir\'e excitons and for their technological application, a  microscopic understanding of underlying bosonic interactions is of key importance. In particular, a theoretical description of the exciton-exciton interaction resulting from their fermionic substructure \cite{combescot2007exciton, katsch2018theory}, and the interplay with the superlattice moir\'e potential has so far remained challenging. Recent experimental studies have demonstrated that the mobility of moir\'e excitons strongly depends on the particle density \cite{choi2020moire,wang2021diffusivity}. However, the interaction between moiré excitons and the effect of elevated densities on  their localization has not been well studied so far. 

In this work, we predict that the lattice of moir\'e-trapped interlayer excitons strongly delocalizes across the moir\'e potential (Fig. \ref{fig:scheme}d) already at intermediate densities ($10^{11}$cm$^{-2}$-$10^{12}$cm$^{-2}$) depending on twist angle and temperature. This interaction-induced delocalization is accompanied by a collapse of the characteristic series of optical moir\'e resonances into a single free exciton peak. Our simulations are based on a novel theoretical framework in which we apply the equilibrium Greens function formalism to derive an extended version of the Gross-Pitaevskii equation for a system of interacting excitons within an external potential. The material-specific parameters are obtained from first principles calculations, which allows us to make material-realistic simulations and predict experimentally accessible key quantities, such as critical densities and characteristic moir\'e features in optical spectra. Moreover, we demonstrate how the interaction-induced delocalization leads to density-tunable supercell tunneling which is crucial for exciton transport and can enable novel technological concepts.   \\ 

{\bf{Results}}

In this work, we focus on the configuration of charge carriers in quasi-equilibrium, i.e. after the initial formation and thermalization of interlayer excitons. To this end, we apply the Greens function formalism to determine the thermal quantum statistics of particles occupying the energetically lowest states at the K point of their respective layer (Fig. \ref{fig:scheme}a). Here, we set up a many-particle Hamiltonian parameterized via first principles calculations for the electronic band structure \cite{kormanyos2015k}, the twist-angle dependent moir\'e potential \cite{brem2020tunable} and the dielectric constants \cite{laturia2018dielectric}, governing the strength of intra- and interlayer Coulomb interactions \cite{ovesen2019interlayer,merkl2019ultrafast}. Based on the equation of motion approach, we derive a coupled system of equations for the electron spectral function and the excitonic eigenstates \cite{steinhoff2017exciton, semkat2009ionization} and show how a self-consistent treatment results in a non-linear eigenvalue problem for excitons, representing a generalized form of the Gross-Pitaevskii equation for bosonic condensates. In particular, we find that the exciton center-of-mass (CoM) motion is modulated by an interplay of the periodic moir\'e potential and the repulsive dipolar repulsion between interlayer excitons. We solve the non-linear eigenvalue problem numerically and deduce optical properties of the system via the luminescence Bloch equations\cite{kira1997quantum, kira2011semiconductor}. Further details are provided in the Supplementary Information.

As start of our analysis, we compare the low-density limit, where interaction effects are expected to be negligible, with the case of a high exciton density. Figure \ref{fig:loc-deloc} summarizes the properties of interlayer moir\'e excitons in hBN-encapsulated MoSe$_2$/WSe$_2$ with a $1^\circ$ twist angle and at a cryogenic temperature ($T=4K$) comparing a low ($n_X=10^8$cm$^{-2}$) and a high density case ($n_X=10^{12}$cm$^{-2}$). The mini-band structure of moir\'e excitons reveals their energy as function of the (quasi) CoM momentum and is displayed on a straight path from a corner of the hexagonal mini-Brillouin zone ($-\kappa$) to the opposite site ($\kappa$) (Fig. \ref{fig:loc-deloc}a and b). The-low density limit resembles the system studied previously \cite{brem2020tunable}, where the moir\'e potential leads to a trapping of exciton states reflected by a characteristic series of flat exciton bands (Fig. \ref{fig:loc-deloc}a) each corresponding to an array of localized excitonic orbitals. Here, only higher-order exciton bands with sufficient kinetic energy can escape the deep potential wells and delocalize across the moir\'e pattern. The latter is reflected by the increased band curvature of the higher-order bands corresponding to a non-zero group velocity. When the exciton density is increased  (Fig. \ref{fig:loc-deloc}b), we find a dramatic change in the band structure: The flat bands have disappeared and instead we find dispersed bands that cover most energies except for small band gaps of $\sim 3$meV at intersection points of different branches.  These curved bands indicate that excitons are no longer trapped but can move through the superlattice with an altered effective mass. This transition from trapped states at low densities to delocalized waves at high densities is also demonstrated by the real space density distributions in Figs. \ref{fig:loc-deloc}c, d. Whereas the trapped states appear as a disconnected array of sharp peaks, the delocalized states have much broader density modulations with non-zero probabilities in between the potential minima. The clear triangular shape of the peaks in Fig. \ref{fig:loc-deloc}d reflects the geometry of the moir\'e potential \cite{brem2020tunable}.

The microscopic origin of this fundamental change in exciton properties at high densities is their bosonic character. Due to the absent Pauli blocking experienced by fermions, bosons tend to bunch in the energetically lowest state. At the chosen high density limit ($n_X=10^{12}$cm$^{-2}$), each supercell is occupied by around three excitons ($\nu=n_XA_{UC}\approx3$). Without including the exciton-exciton interaction, these particles would all occupy the lowest band in Fig. \ref{fig:loc-deloc}a corresponding to the sharp density pockets in Fig. \ref{fig:loc-deloc}c. However, the resulting large local density of excitons leads to a significant electrostatic repulsion due to the dipolar nature of interlayer excitons. In particular, the repulsive force between the excitons gives rise to a mean-field Coulomb potential that is comparable in strength with the moir\'e potential. As a result, the effective total potential is significantly shallower than the original bare moir\'e potential (Fig. \ref{fig:scheme}d).  This in turn gives rise to a broader density distribution, which leads to a weaker repulsion. This interplay between moir\'e trapping and repulsion-driven expansion is computed iteratively until a self-consistent solution is found. 

It is important to note that the predicted effect is fundamentally different from the saturation effect for localized electron states, such as quantum dots. For the latter, only a single electron can occupy the localized state, such that additional electrons are simply pushed to higher energy shells, leaving the ground state mostly unaffected. In contrast, in the case of bosons, most particles occupy the same (ground) state. Therefore a localized bosonic state can become unstable at high density because the repulsive potential - resulting from multiple excitons in the same state - can push the excitons out of the trap. This collective effect is similar to the interaction-induced expansion of a trapped Bose-Einstein condensate \cite{foot2004atomic}, but has not been discussed in the context of moir\'e excitons.  

Finally, we investigate how the density-dependent moir\'e exciton transitions described above can be detected in optical experiments. In Figs. \ref{fig:loc-deloc}e and f we show the optical absorption spectrum in the low and high density case, respectively. At low densities, the array of trapped states is reflected by a characteristic series of sharp resonances (Fig. \ref{fig:loc-deloc}e), each stemming from a different moir\'e mini-band. In general, only excitons with zero CoM momentum can be directly excited by light, due to the negligible momentum provided by a photon. In spatially homogeneous system, this momentum selection rule is reflected by a single exciton resonance at the energy of the exciton with zero CoM momentum. However, the moir\'e potential leads to a periodic modulation of the exciton wave function corresponding to a superposition of different momenta. Consequently, each moir\'e mini-band obtains a finite oscillator strength according to the probability amplitude at zero momentum. In contrast, at high densities the effective potential is much shallower leading to smaller energy separations and a less pronounced spatial modulation/mixing of momenta. Therefore, at high densities only the lowest subbands appear and they lie closer in energy  (Fig. \ref{fig:loc-deloc}f), closely resembling the single exciton peak in spatially homogeneous systems (i.e. for delocalized  excitons). 

\begin{figure*}[t!]
\includegraphics[width=\textwidth]{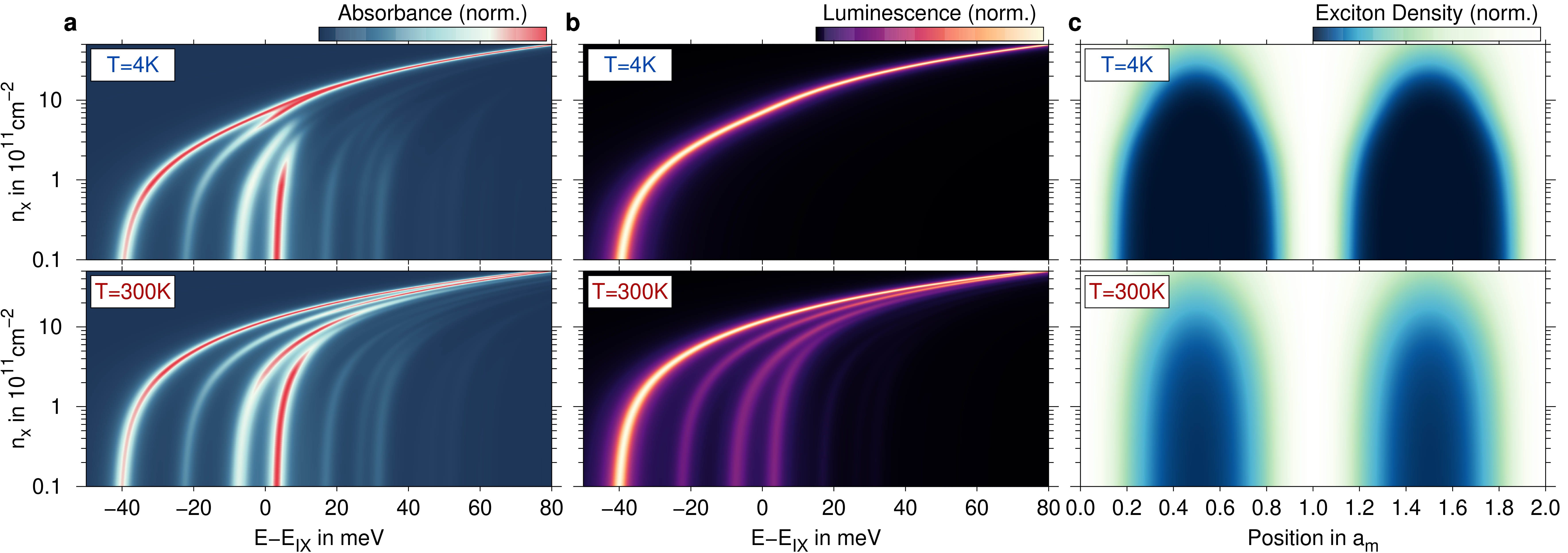}
\caption{{\bf{Density-dependent absorption and emission spectra of moir\'e excitons.}} {\bf{a}}, Absorbance spectra as continuous function of exciton density assuming thermalized distributions with excitonic temperatures of $T=4$K (top) and $T=300$K (bottom). The interaction-induced delocalization gives rise to blue shifts and a decrease of the splitting between different moir\'e resonances until all peaks merge into a single resonance. {\bf{b}}, Corresponding emission spectra (PL) displaying the same general behaviour, however only exhibiting peaks when the corresponding state is occupied by the thermally distributed exciton population. {\bf{c}}, Spatial profile of the exciton density illustrating the delocalization at increased densities. At $T=300$K the localization is softened, which results in weaker Coulomb interactions.}
\label{fig:sweep} 
\end{figure*}

To further understand the interaction-induced delocalization, we analyse the optical spectrum as continuous function of the density and determine critical densities at which interactions become important. Moreover, we compare the absorption features with experimentally more easily accessible photoluminescence (PL) spectra at different temperatures (Fig. \ref{fig:sweep}). First, we again focus on the low-temperature absorption (top panel Fig. \ref{fig:sweep}a). At low densities of $<10^{11}$cm$^{-2}$, we retrieve the same multi-peak spectrum as in Fig. \ref{fig:loc-deloc}e. However, starting at densities of about $10^{11}$cm$^{-2}$ the lowest moir\'e resonance undergoes a strong blue-shift reflecting the onset of repulsive interaction between the localized excitons. At the same time, the spectral distance between different moir\'e resonances (reflecting the depth of the effective total potential) decreases until they merge into a single dominant peak at about $10^{12}$cm$^{-2}$. The rather low critical density of $10^{11}$cm$^{-2}$ for the onset of interactions is owed to the sharp density modulation for the localized states, leading to much higher local densities in the low energy pockets of the moir\'e pattern.

The predicted spectral modifications at increased exciton densities are direct indicators of the interaction-induced delocalization. However, due to their spatially indirect character, the absorption amplitude of interlayer excitons is comparably small and difficult to detect in experiments. Therefore, moir\'e resonances are usually observed in PL measurements. When considering the PL spectrum at low temperatures (top panel Fig. \ref{fig:sweep}b), only the lowest moir\'e peak appears since higher-order states are unoccupied. This is due to the large separation of $\sim20$meV between the exciton states  compared to the thermal energy of $<1$meV at the considered temperature of 4K. Nonetheless, multiple moir\'e peaks with comparable energy separations have been observed in PL experiments at cryogenic temperatures \cite{tran2019evidence}. This indicates that the continuous wave excitation in standard PL experiments gives rise to a quasi-equilibrium distribution of excitons with a significantly larger temperature compared to the cooled atomic lattice. Therefore, we increase the exciton temperature to investigate the impact of a broader energy distribution. The bottom panels of Fig. \ref{fig:sweep}a and b show the respective absorption and emission spectra at an exemplary exciton temperature of $T=300$K. The density dependence of the absorption shows a similar behavior as at 4K, with slight quantitative differences. In particular, the blue-shift of the lowest moir\'e peak becomes efficient at slightly larger densities and the shrinking of energy separations is slower than at 4K. This softening of Coulomb effects is directly related to the broader energy range occupied by excitons at increased temperatures. The bottom panel of Fig. \ref{fig:sweep}b now contains several PL peaks, which is owed to the non-zero occupation of higher energy mini-bands. These states correspond to localized orbitals with larger kinetic energy and therefore increased spatial extent. Consequently, the overall exciton density is less tightly concentrated, which is illustrated in the density profiles in Fig. \ref{fig:sweep}c. As a result of the enhanced kinetic energy at high temperatures, the exciton distribution is localized less tightly. Therefore, the interaction-induced delocalization occurs at larger mean densities, since the modulation of the local density is softened, i.e. excitons are distributed over larger spatial areas. Our simulation shows that the efficiency of interaction-induced effects is strongly connected to the spatial modulation of the exciton density, which e.g. can be controlled by temperature. Another important knob to control the spatial arrangement of moir\'e excitons is the twist angle of the heterostructure, which will be investigated next.

\begin{figure}[t!]
\includegraphics[width=\columnwidth]{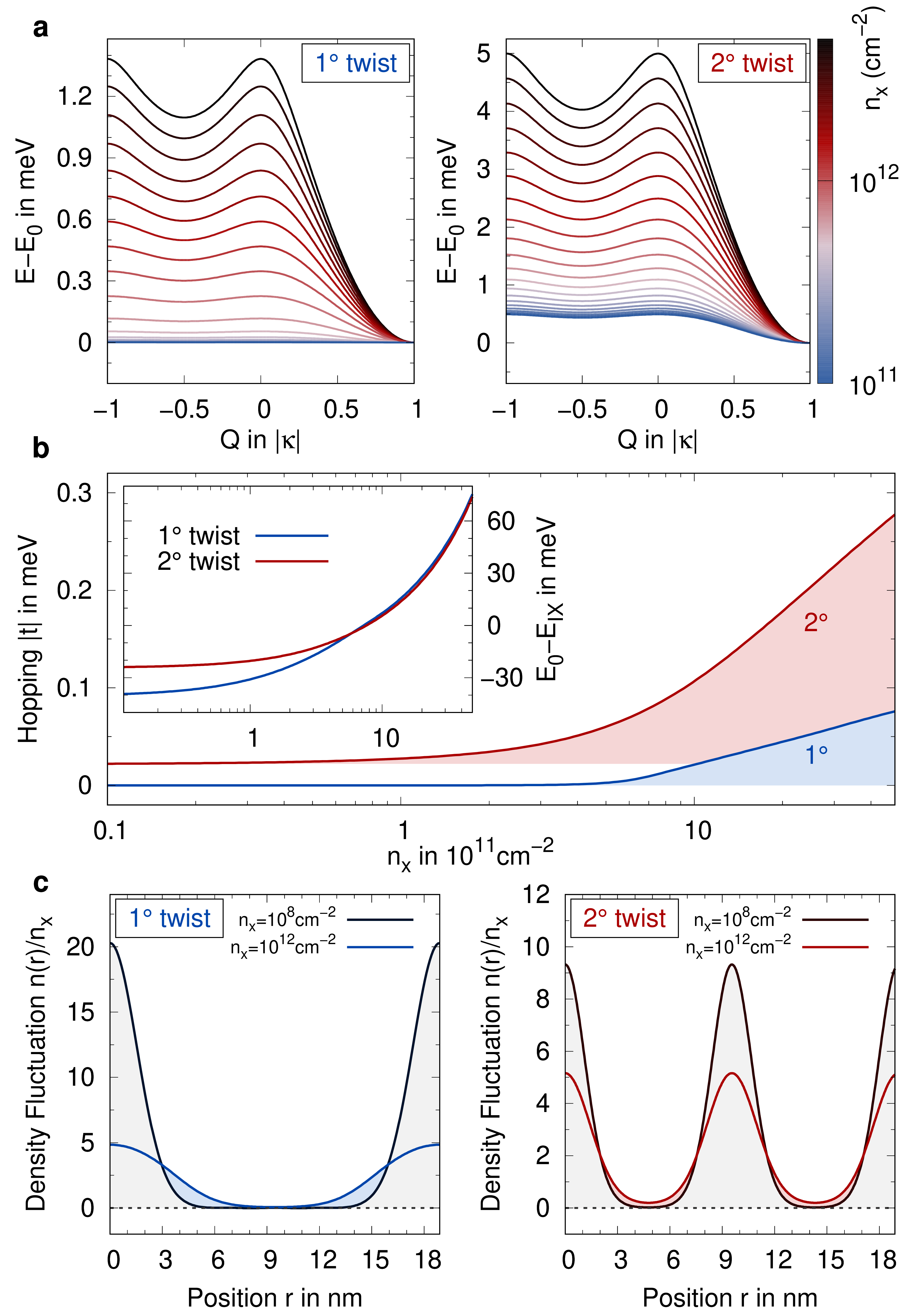}
\caption{{\bf{Density-tunability of moir\'e excitons at different twist angles.}} {\bf{a}}, Evolution of the lowest moir\'e mini-band at different densities at $T=4K$ and the 1$^\circ$ (left) or 2$^\circ$ twist angle (right). Whereas at 1$^\circ$ the low density limit exhibits a completely flat band, the different confinement condition at 2$^\circ$ results in non-zero bandwidth and a stronger impact of interaction effects. {\bf{b}}, Band modifications are quantified via the next-neighbour hopping energies and energy shifts (inset) as function of density. {\bf{c}}, Real space density profiles at low and high density for two different twist angles. The different density compression and intercell distance leads to stronger interactions at 1$^\circ$, but a more significant change in the hopping rates at 2$^\circ$.   }
\label{fig:hopping} 
\end{figure}

When the twist angle is increased from $\theta=1^\circ$ to 2$^\circ$, the period of the moir\'e pattern $a_\text{m}=a_0/\sqrt{2(1-\cos \theta)}\approx a_0/\theta$ is decreased by a factor of 1/2, while the depth of the potential is approximately constant. This direct decrease in the confinement length leads to an increase of the kinetic energies and a larger spatial extent of the localized states with respect to the supercell size. Therefore, intercell hopping is naturally enhanced when the twist angle is increased \cite{brem2020tunable}. In the following, we investigate how the intercell exciton transport is further modified when interactions effects become important. In Figure \ref{fig:hopping}, we compare the density-dependent transport properties at the twist angles of 1$^\circ$ and 2$^\circ$. 
First, we consider the curvature of the lowest moir\'e mini-bands in Fig. \ref{fig:hopping}a. For 1$^\circ$ twist and low densities we retrieve the completely flat band of localized excitons already discussed in Fig. \ref{fig:loc-deloc}a. When the density is gradually increased, the bands become broader and more dispersed, reflecting the interaction-induced delocalization due to Coulomb-driven expansion. The same effect is also observed at 2$^\circ$ twist, with the difference that the low-density band already exhibits a significant curvature as result of the above described change in the confinement conditions. Moreover, the change in the band curvature is more significant at 2$^\circ$ for the same density indicating an enhanced interaction effect. To quantify the band modifications and their impact on transport, we consider the next-neighbour hopping strength (Fig. \ref{fig:hopping}b), which is directly connected to the band width and curvature. Moreover, we also investigate the blue-shift of the lowest mini-band as function of density (inset). We find that the overall change in the hopping strength is significantly more pronounced for 2$^\circ$ twist and already starts at much smaller densities. In contrast, the blue-shift of the exciton energy starts at lower densities for 1$^\circ$ than for 2$^\circ$ (inset) indicating stronger interactions for lower twist angles. This apparent contradiction can be microscopically explained by comparing the spatial modulation of the density at 1$^\circ$ and 2$^\circ$ and how it changes when the number of excitons is increased (Fig. \ref{fig:hopping}c).

At low densities and  1$^\circ$ twist, we find strongly localized states with a small next-neighbour overlap. As a consequence of the moir\'e potential, excitons are trapped in low-energy pockets leading to an increase of the local density by a factor of 20 with respect to the mean density (left panel in Fig. \ref{fig:hopping}c). For the larger twist angle, the moir\'e cells are more closely packed, such that the density is distributed to more sites, i.e. for a fixed number of excitons the filling factor is smaller at 2$^\circ$ than at 1$^\circ$. The latter results in a less effective compression of the density with peak values $< 10$ (right panel in Fig. \ref{fig:hopping}c). This already explains the earlier onset of the blue-shift for 1$^\circ$ in the inset of Fig. \ref{fig:hopping}b. As result of the stronger compression, the local density within low energy pockets is much higher at 1$^\circ$ resulting in stronger interactions. Nevertheless, the next-neighbour hopping $t$, which reflects the wave function overlap between neighbouring cells, is more strongly affected by the interaction-induced expansion at larger twist angles. This results from the fact that the localization sites are more closely spaced for larger twist angles, such that a smaller net change of the orbital width is necessary to significantly change the next neighbour overlap. In summary, interactions and peak shifts are stronger at smaller twist angles, but interestingly the transport properties are more effectively modified at larger twist angles, since changes in the orbital width become more important, the closer the localization sites are. \\

{\bf{Discussion}}

Our study demonstrates the importance of interaction effects for the properties of interlayer excitons in moir\'e potentials. Moreover, our findings can be translated to other systems of interacting bosons being trapped in external potentials, such as ultracold atoms trapped in optical lattices. Most importantly, the predicted interaction-induced delocalization and the connected changes in optical spectra are a natural consequence of the bosonic nature of moir\'e excitons. In particular, the predicted blue-shift and the decrease of moir\'e peak splittings can only occur for trapped bosons whose ground state can be occupied by several particles. This is fundamentally different from the well-known saturation effect occurring in systems of localized electrons. For fermions the emission energies of the localized emitters do not shift with excitation power, but the peak intensity saturates when each site is filled with one particle. Therefore, the predicted density-dependence of emission features can serve as an experimental benchmark to distinguish genuine moir\'e exciton features, i.e. bosons trapped in periodic long-range potentials, from other trapping mechanisms, such as the localization of electrons at lattice defects. 
Finally, we predict that the interaction-induced expansion of localized states leads to a density-dependent hopping rate. The connected change in the exciton mobility, when going from an array of trapped states at low densities to a liquid of delocalized excitons at $n_x \sim 10^{11}$cm$^{-2}$, has recently been observed in diffusion experiments \cite{wang2021diffusivity}. Such a strong non-linearity in the exciton transport properties can be utilized in technological applications. In particular, the delocalization at high density opens up the possibility to control the strength of excitonic currents via adjustment of the chemical potential, e.g. through illumination of the material, i.e. optical generation of excitons. In this way excitonic interactions in moir\'e materials could enable novel technological concepts in combined photonic and excitonic circuits \cite{high2008control}. 

To summarize, we have developed a microscopic theory that allows us to study the Coulomb interaction between correlated electron-hole pairs within external potentials. We find that the strong repulsion between interlayer excitons combined with their bosonic nature gives rise to a delocalization of moir\'e excitons in twisted MoSe$_2$/WSe$_2$ heterostructures at experimentally relevant densities of $10^{11}-10^{12}$cm$^{-2}$. This interaction-induced delocalization is accompanied by a blue-shift and a decrease of the splitting between optical moir\'e resonances. Furthermore, the underlying mechanism, i.e. repulsion-driven expansion of moir\'e orbitals, has a significant impact on the intercell hopping rates. This can be utilized to tune exciton transport properties, whereas the twist-angle can be used to control the critical density at which interaction effects become significant. Our work provides new theoretical tools to study strongly correlated matter in quantum materials and predicts concrete experimentally accessible features in optical spectra. Our predictions will help to interpret existing studies and will guide future experiments on moir\'e excitons and other bosonic systems.   \\

\begin{acknowledgments}
We acknowledge support from Deutsche Forschungsgemeinschaft (DFG) via SFB 1083 (Project B9) and the European Unions Horizon 2020 research and innovation program under grant agreement No 881603 (Graphene Flagship).
\end{acknowledgments}

\bibliographystyle{achemso}

\section*{References}
\providecommand{\latin}[1]{#1}
\makeatletter
\providecommand{\doi}
{\begingroup\let\do\@makeother\dospecials
	\catcode`\{=1 \catcode`\}=2 \doi@aux}
\providecommand{\doi@aux}[1]{\endgroup\texttt{#1}}
\makeatother
\providecommand*\mcitethebibliography{\thebibliography}
\csname @ifundefined\endcsname{endmcitethebibliography}
{\let\endmcitethebibliography\endthebibliography}{}

\end{document}